# Superstatistics, thermodynamics, and fluctuations

Sumiyoshi Abe[1,2], Christian Beck[3], and Ezechiel G. D. Cohen[4]

[1]*Department of Physical Engineering, Mie University, Tsu, Mie 514-8507, Japan*
[2]*Institut Supérieur des Matériaux et Mécaniques Avancés, 44 F. A. Bartholdi, 72000 Le Mans, France*
[3]*School of Mathematical Sciences, Queen Mary, University of London, Mile End Road, London E1 4NS, UK*
[4]*The Rockefeller University, 1230 York Avenue, New York, New York 10021, USA*

**Abstract**   A thermodynamic-like formalism is developed for superstatistical systems based on conditional entropies. This theory takes into account large-scale variations of intensive variables of systems in nonequilibrium stationary states. Ordinary thermodynamics is recovered as a special case of the present theory, and corrections to it can systematically be evaluated. A generalization of Einstein's relation for fluctuations is presented using a maximum entropy condition.

## I. INTRODUCTION

Nonequilibrium complex systems often exhibit dynamics that can be decomposed into several dynamics on different time scales. As a simple example, consider a Brownian particle moving through a changing fluid environment, characterized by temperature variations on a large scale. In this case, two dynamics are relevant: one is a fast dynamics describing the local motion of the Brownian particle and the other one is a slow one due to the large global variations of the environment with spatio-temporal inhomogeneities. These effects produce a superposition of two different statistics, which is referred to as superstatistics.

The concept of superstatistics has been introduced by two of the present authors [1] after some preliminary considerations in [2,3]. The stationary distributions of superstatistical systems typically exhibit a non-Gaussian behavior with fat tails, which can decay, e.g., as a power law, a stretched-exponential law, or in an even more complicated way [4]. Essential for this approach is the existence of an intensive variable $\beta$, which fluctuates on a large spatio-temporal scale.

For the above-mentioned example of a superstatistical Brownian particle, $\beta$ is the fluctuating inverse temperature of the environment. In general, however, $\beta$ may also be an effective friction constant, a changing mass parameter, a variable noise strength, the fluctuating energy dissipation in turbulent flows, a fluctuating volatility in finance, an environmental parameter for biological systems, a local variance parameter extracted from a signal, and so on.

Superstatistics offers a very general framework for treating nonequilibrium stationary states of such complex systems. After the original work in Ref. [1], a lot of efforts have been made for further theoretical elaboration [5-12]. At the same time, it has also been applied successfully to a variety of systems and phenomena, including hydrodynamic turbulence [9,13,14], pattern formation [15], cosmic rays [16], solar flares [17], mathematical finance [18-20], random matrices [21], complex networks [22], wind velocity fluctuations [23], and hydro-climatic fluctuations [24].

Due to these successes, it appears meaningful now to study the macroscopic properties of superstatistical systems. Thus, the purpose of this paper is not to examine further applications of superstatistics but to address the following issue: Is it possible to establish an effective thermodynamic-like macroscopic formalism for superstatistics? For this purpose, we will consider conditional entropies associated with local temperature variations, which allow us to develop a consistent formalism. We will also show that ordinary equilibrium thermodynamics is recovered as a special case when there are no temperature variations, and we will calculate systematic corrections to ordinary thermodynamics by analytically treating a sharply-peaked distribution of the temperature variations. In addition, a conditional entropy turns out to generalize Einstein's theory of fluctuations [25] in conformity with a maximum entropy condition without any *a priori* constraints.

This paper is organized as follows. In Sec. II, conditional entropies are introduced and a thermodynamic-like formalism is developed for superstatistics. In Sec. III, a superstatistical correction to ordinary thermodynamics is systematically evaluated. In Sec. IV, the temperature variations are described by making use of the maximum entropy condition, and the result can be interpreted as a generalization of Einstein's theory of fluctuations. An application of the theory to superstatistical Brownian particles is discussed in Sec. V. Sec. VI is devoted to concluding remarks. For the sake of simplicity, the Boltzmann constant is set equal to unity throughout the paper.

## II.   CONDITIONAL ENTROPY AND THERMODYNAMIC FORMALISM FOR SUPERSTATISTICS

Let us first recall the basic idea underlying superstatistics. We will then proceed to the definition of a conditional entropy function and the formulation of the associated thermodynamic formalism.

Consider a complex system in a nonequilibrium stationary state that is driven by some external forces. Such a system will be, in general, inhomogeneous in both space and time. Effectively, it may be thought to consist of many spatial cells (or, the time series may consist of many time slices), in each of which there may be a different value of some relevant system parameter, $\beta$. Its inverse, $\beta^{-1}$, is a local variance parameter of a suitable observable of the complex system. The cell size is effectively determined by the condition that it is small compared to the correlation length of the $\beta$-field as measured on a large scale. A superstatistical system is characterized additionally by the condition that the local relaxation time of the system is short compared to the typical time scale of changes of $\beta$, so that each cell can be formally assumed to be in local equilibrium. Sometimes this property will be satisfied for a given complex system, sometimes not [9]. For our approach to be applicable, we must have sufficiently large separation of these two time scales. (From the above, it is clear that superstatistics is a nonequilibrium concept and has nothing to do with the estimator approach of Ref. [26].)

It should also be clear that the meaning of the mathematical variables is different in various applications to complex systems. So, "local equilibrium" is meant in a generalized sense for suitable observables of the system dynamics under consideration. In the long term, the stationary distribution of a superstatistical inhomogeneous system arises as superposition of a local Boltzmann factor $e^{-\beta E}$ (or analogues of the Boltzmann factor) with various values of $\beta$ weighted with a global probability density $f(\beta)$ to observe some value $\beta$ in a randomly chosen cell:

$$p(E) = \int d\beta \, f(\beta) \frac{1}{Z(\beta)} \rho(E) \, e^{-\beta E}. \qquad (1)$$

Here, $E$ is an effective energy associated with each cell, $Z(\beta)$ the normalization constant of $\rho(E) e^{-\beta E}$ for a given $\beta$, and $\rho(E)$ the density of energy states. Clearly, while the energy $E$ is well defined for simple physical systems, it will be an effective physical parameter in general, so that $e^{-\beta E}$ describes the local equilibrium distribution of a suitable observable, $E$, in each cell. For example, if there is locally Gaussian behavior of a suitable observable (e.g., a velocity $\mathbf{v}$) in each spatial cell, then the effective local Hamiltonian contains just the "kinetic energy", $E = \mathbf{v}^2 / 2$, of a particle with unit mass. The long-term stochastic process then consists of a superposition of Gaussian factors with a fluctuating variance $\beta^{-1}$.

As stated before, our aim here is to develop a thermodynamic-like formalism that is applicable to a wide class of complex systems with large separations of time scales. For this purpose we introduce a conditional Boltzmann-Gibbs entropy for a superstatistical nonequilibrium system by taking into account the fluctuating $\beta$. This conditional entropic measure and the corresponding thermodynamics developed from it have a direct physical interpretation and it differs radically from previous work in Ref. [6], which introduces a generalized entropy. We believe that our approach is physically relevant in the sense that the thermodynamic-like relations obtained here correctly describe the physics of superstatistical nonequilibrium systems. We next introduce the conditional entropy.

Consider first, in general, two random variables $X$ and $Y$, which are not necessarily independent of each other. (Later, in the superstatistical application, $X$ will correspond to the energy and $Y$ to the inverse temperature of a cell, but, for the moment, we just restrict ourselves to general arguments.) The possible outcomes (events or microstates, for example) of $X$, $X_i$, are labeled by the index $i$ and those of $Y$, $Y_j$, by $j$, respectively. The joint probability of the event $(i, j)$ is denoted by $p_{ij}(X, Y)$. Let us look at the Boltzmann-Gibbs entropy $S[X, Y]$ associated with the joint system

$$S[X, Y] = -\sum_{i,j} p_{ij}(X, Y) \ln p_{ij}(X, Y). \tag{2}$$

Bayes' rule states that $p_{ij}(X, Y) = p_{ij}(X | Y) p_j(Y) = p_{ij}(Y | X) p_i(X)$, where $p_{ij}(X | Y)$ is the conditional probability that event $i$ takes place if we already know that the event $j$ has happened, and $p_j(Y) = \sum_i p_{ij}(X, Y)$ is the marginal probability. Substituting this relation into Eq. (2), we immediately obtain

$$S[X, Y] = S[X | Y] + S[Y] = S[Y | X] + S[X] \tag{3}$$

where $S[X | Y]$ is the conditional entropy defined by

$$S[X|Y] = \sum_j S[X|Y_j) p_j(Y) \tag{4}$$

with

$$S[X|Y_j) = -\sum_i p_{ij}(X|Y) \ln p_{ij}(X|Y), \tag{5}$$

which is a function only of $Y_j$, since one has summed over the $X_i$.

Let us now apply these general considerations to a superstatistical system. In this case, let $X$ correspond to the energy $E$ in a given spatial cell (we assume that ordinary equilibrium statistical mechanics with the energy $E$ is locally valid), and $Y$ to an additional random variable describing the fluctuating inverse temperature in the various spatial cells. From now on this additional random variable will be denoted by $B$. Thus, we obtain

$$S[E, B] = S[E|B] + S[B]$$
$$= \int d\beta\, f(\beta) S[E|\beta) - \int d\beta\, f(\beta) \ln f(\beta), \tag{6}$$

where we have replaced the sums over $j$ by integrals over $\beta$. In the local cells, the conditional probability $p(E = \varepsilon_i | \beta)$ to observe the microstate $i$ with the energy $\varepsilon_i$ is given by the canonical ensemble with the inverse cell temperature $\beta$:

$$p(\varepsilon_i | \beta) = \frac{1}{Z(\beta)} e^{-\beta \varepsilon_i}, \tag{7}$$

where $Z(\beta)$ is the canonical partition function and, from this, $S(E|\beta)$ in the integrand on the right-hand side of Eq. (6) is given by $S(E|\beta) = -\sum_i p(\varepsilon_i|\beta) \ln p(\varepsilon_i|\beta)$, which is a function of $\beta$ only, since one has summed over the energies. Substituting Eq. (7) into Eq. (6), we arrive at the basic result

$$S[E, B] = \overline{\beta U(\beta) + \ln Z(\beta)} + S[B], \tag{8}$$

where we denote the average of an arbitrary observable $Q$ over the fluctuating inverse temperatures as $\overline{Q(\beta)} \equiv \int d\beta \, f(\beta) Q(\beta)$ and $U(\beta) = \sum_i \varepsilon_i \, p(\varepsilon_i | \beta)$ is the internal energy in each cell. [In the discrete notation of Eq. (7), the density of states is omitted for the sake of simplicity.]

The entropy $S[E, B]$ has contributions from both ordinary equilibrium states with $\varepsilon_i$'s in the local cells and the distribution of the global temperature variations. The separation into two scales is explicitly implemented here by the use of conditional concepts. In fact, *the randomness of $\beta$ is quenched in $U(\beta)$ and $\ln Z(\beta)$, and averaging over $\beta$ is performed afterwards*. This is in marked contrast to the previous work in Ref. [6], which introduced a generalized entropy and does not explicitly describe the existence of two scales in a superstatistical system.

Clearly, if there are no temperature variations at all, i.e., $f(\beta) = \delta(\beta - \beta_0)$, we have, after appropriate regularization, $S[B] = 0$ as well as $\overline{\beta U(\beta)} = \beta_0 U(\beta_0)$ and $\overline{\ln Z(\beta)} = \ln Z(\beta_0)$. Therefore, in such a special case, we obtain

$$F = -\beta_0^{-1} \ln Z(\beta_0) = U(\beta_0) - \beta_0^{-1} S[E, \beta_0], \tag{9}$$

i.e., the ordinary expression for the equilibrium Helmholtz free energy at inverse temperature $\beta_0$.

### III. SHARPLY PEAKED TEMPERATURE VARIATIONS

Let us now study which type of thermodynamics is generated by the entropy in Eq.

(8) if the inverse temperature variations are sharply peaked around an average value $\beta_0$. Here, we are particularly interested in the term $\overline{\beta U(\beta)}$. We may write

$$\overline{\beta U(\beta)} = c \sum_i \int d\beta \, \tilde{f}(\beta) \, \varepsilon_i \, e^{-\beta \varepsilon_i}, \qquad (10)$$

where

$$\tilde{f}(\beta) = \frac{\beta}{c} \frac{f(\beta)}{Z(\beta)} \qquad (11)$$

is yet another normalized probability distribution with the constant $c$ determined by

$$c = \int d\beta \, \beta \frac{f(\beta)}{Z(\beta)}, \qquad (12)$$

which has dimension $(\text{energy})^{-1}$. Eq. (10) is further rewritten as follows:

$$\overline{\beta U(\beta)} = c \sum_i \varepsilon_i \, \text{B}(\varepsilon_i), \qquad (13)$$

where $\text{B}(\varepsilon_i)$ is the generalized Boltzmann factor [1]:

$$\text{B}(\varepsilon_i) = \int d\beta \, \tilde{f}(\beta) \, e^{-\beta \varepsilon_i}. \qquad (14)$$

If $f(\beta)$ is sharply peaked, then so is $\tilde{f}(\beta)$. Following Ref. [1], we can expand the generalized Boltzmann factor for a peaked distribution as

$$\Beta(\varepsilon_i) = e^{-\beta_0 \varepsilon_i} \left(1 + \frac{1}{2}\sigma^2 \varepsilon_i^2 + \cdots \right), \tag{15}$$

where $\sigma^2$ is the variance of inverse temperature fluctuations calculated with the distribution function $\tilde{f}(\beta)$, that is, $\sigma^2 = \int d\beta\, \beta^2\, \tilde{f}(\beta) - \left(\int d\beta\, \beta\, \tilde{f}(\beta)\right)^2$. Thus, Eq. (13) is evaluated as follows:

$$\overline{\beta U(\beta)} = c \sum_i \varepsilon_i\, e^{-\beta_0 \varepsilon_i} \left(1 + \frac{1}{2}\sigma^2 \varepsilon_i^2 + \cdots \right)$$

$$= c \langle E \rangle + \frac{c}{2}\sigma^2 \langle E^3 \rangle + \cdots, \tag{16}$$

where we have introduced a notation for "unnormalized" canonical averages, $\langle E^m \rangle = \sum_i \varepsilon_i^m\, e^{-\beta_0 \varepsilon_i}$. This result is a kind of modified thermodynamic-like expression for a superstatistical system. It is based on canonical averages $\langle E^m \rangle$ of the above type with a fixed average inverse temperature $\beta_0$. However, in Eq. (16), the average energy $\langle E \rangle$ is not multiplied by $\beta_0$, but rather by $c$, which is close to $\beta_0$, since $f(\beta)$ is sharply peaked. Moreover, there is a leading-order correction term proportional to the variance $\sigma^2$ of the temperature variations combined with the canonical average of the third power of the energy.

The above consideration shows that, in leading order of the moments of the energy, it is possible to reduce a superstatistical thermodynamics, generated by the entropy in Eq. (8), to ordinary thermodynamics with slightly different effective energy and slightly different types of averages.

It is noted, however, that there are certain situations in which $f$ may not have peaks. We shall discuss this point in the following two sections.

## IV. DISTRIBUTION OF TEMPERATURE VARIATIONS AND SUPERSTATISTICAL GENERALIZATION OF EINSTEIN'S FLUCTUATION RELATION

Now we pose the following question. Suppose we have a complex system described by superstatistics. Is there a principle for determining the distribution of the large-scale temperature variations in the system? The answer to this question depends on the physical situation, i.e., how much information is available about the system. Therefore, it seems natural to apply a condition of maximum entropy under certain constraints.

The physical situation we consider here is the simplest one, in which no *a priori* information is available [27]. Accordingly, the entropy in Eq. (6) is conditionally maximized under the constraint of normalization of $f(\beta)$ only. That is,

$$\delta_f \left\{ S[E,B] - \alpha \left( \int d\beta \, f(\beta) - 1 \right) \right\} = 0, \tag{17}$$

where $\alpha$ is a Lagrange multiplier. Recall that the short time scale of the dynamics was already averaged out, and thus $S[E, B]$ can be regarded as a functional of $f(\beta)$ only. The solution of this problem is given by

$$f(\beta) = \text{const} \cdot e^{S[E|\beta]}. \tag{18}$$

with

$$S[E \mid \beta) = \beta U(\beta) + \ln Z(\beta), \tag{19}$$

where $S[E \mid \beta)$ is a function of $\beta$ only, while $E$ is nothing but a dummy variable, indicating merely the nature of the conditional probability [cf. Eq. (5)]. At this stage, we see a striking similarity between Eq. (18) and Einstein's theory of fluctuations [25], which was an inversion of Boltzmann's $S = \log W$ ($k \equiv 1$) to $W \sim e^S$, where $S$ is the thermodynamic entropy of the system under consideration. It should be noted, however, that the entropy appearing in Eq. (18) is not the entropy itself but a conditional entropy, conditioned by the quenched temperature fluctuations. It should also be noted that we are concerned with fluctuations in a nonequilibrium system. In fact, Eq. (17) describes a procedure of a conditional maximization of $S[E, B]$, not the total maximization characterizing equilibrium. And this is precisely the point in which our discussion deviates from Einstein's theory. On the other hand, Eq. (18) reduces to Einstein's relation if the system is in a state near equilibrium and the temperature variations are small. In this way, Einstein's theory of fluctuations is generalized by using conditional entropies.

Closing this section, we note that $S[E|B)$ tends to decrease in a monotonic way with respect to $\beta$, and accordingly $f(\beta)$ in Eq. (18) may not have peaks, in general. This is another point which differs from Einstein's theory. If some more information is available, we have a further constraint on the average value of a certain quantity, $Q(\beta)$, in the variational principle in Eq. (17). Then, the resulting distribution is given by

$$f(\beta) = \text{const} \cdot e^{S[E|\beta] - \lambda Q(\beta)}, \tag{20}$$

where $\lambda$ is a Lagrange multiplier. Depending on the property of $Q(\beta)$, $f(\beta)$ can have a peak.

## V. A SIMPLE EXAMPLE: MUTUALLY NONINTERACTING SUPERSTATISTICAL PARTICLES

Let us now examine, as an illustration of the foregoing considerations, a simple model of a superstatistical system consisting of $n$ non-interacting classical Brownian particles with unit mass in the spatial cells of a fluid that is subject to large scale temperature variations. Given a local inverse temperature $\beta$ in a given cell, the conditional probability of finding the momenta (i.e., the velocities) $\mathbf{v}_1, \mathbf{v}_2, \cdots, \mathbf{v}_n$ in the cell is given by [cf. Eq. (7)]

$$p(\mathbf{v}_1, \mathbf{v}_2, \cdots, \mathbf{v}_n | \beta) = \frac{1}{Z(\beta)} \exp\left[-\frac{\beta}{2}(\mathbf{v}_1^2 + \mathbf{v}_2^2 + \cdots + \mathbf{v}_n^2)\right] \quad (21)$$

with the local partition function

$$Z(\beta) = \frac{v^n}{n!}\left(\frac{2\pi}{h^2 \beta}\right)^{3n/2}, \quad (22)$$

where $v$ and $h^3$ are the volumes of the spatial cells and those of appropriate cells in phase space, respectively. In addition, the local internal energy is

$$U(\beta) = \frac{3}{2} n \beta^{-1}. \quad (23)$$

Therefore, $S[E|\beta)$ is calculated to be

$$S[E|\beta) = \beta U(\beta) + \ln Z(\beta)$$

$$= n\left(-\frac{3}{2}\ln \beta + c_0\right), \quad (24)$$

where $c_0 = -(3/2)\ln[h^2/(2\pi)] + \ln(v/n) + 5/2$. Thus, the generalized Einstein relation in Eq. (18) yields

$$f(\beta) \sim \beta^{-3n/2}. \quad (25)$$

This is purely a power-law distribution and does not have peaks, as mentioned in the previous section. It is normalizable only over a finite range of $\beta$, $(\beta_{\min}, \beta_{\max})$, where $\beta_{min}$ ($\beta_{max}$) can be small (large) but finite. This situation may be physically plausible if for example the Brownian particles in a turbulent fluid flow are considered, since in such a fluid state finite $\beta_{min}$ and $\beta_{max}$ are expected to exist.

$\tilde{f}(\beta)$ in Eq. (11), which appears in Eq. (14), is then found to be

$$\tilde{f}(\beta) \sim \beta. \tag{26}$$

Now, as shown in Ref. [8], any distribution $\tilde{f}(\beta)$ behaving for small $\beta$ as

$$\tilde{f}(\beta) \sim \beta^{\gamma} \quad (\gamma > 0) \tag{27}$$

implies that the generalized Boltzmann factor in Eq. (14) decays for large values of the energy as

$$B(\varepsilon_i) \sim \varepsilon_i^{-1-\gamma}. \tag{28}$$

Eq. (26) requires $\gamma$ to be

$$\gamma = 1. \tag{29}$$

It may be also of interest to compare Eq. (28) with the asymptotic behavior (i.e., large $\varepsilon_i$) of the statistical factor in Tsallis statistics

$$B(\varepsilon_i) \sim \varepsilon_i^{1/(1-q)}, \tag{30}$$

where $q$ is Tsallis' entropic index [28]. This comparison leads to the following value of the entropic index:

$$q \equiv 1 + \frac{1}{1+\gamma} = \frac{3}{2}. \tag{31}$$

As a matter of fact, this same value of $q$ is also encountered in the description of many experiments on complex systems (e.g., of small-scale hydrodynamic turbulence [3] and of pattern forming systems [15]). This suggests that the present theory, with more structured forms for the conditional entropy than used here, could perhaps be used to understand the typical behavior of complex systems.

Finally, it is also of interest to investigate the case of an additional constraint on the variance of $\ln \beta$, i.e., $Q(\beta) \sim (\ln \beta)^2$, in Eq. (20). In this case, $f(\beta)$ in Eq. (20) has the form of the log-normal distribution, which is now normalizable in the full range of $\beta$. Associated log-normal superstatistics is known to be relevant to, for example, Lagrangian turbulence [29], where variations of $\beta$ describe fluctuations of energy dissipation.

## VI. CONCLUDING REMARKS

We have developed a thermodynamic-like formalism for superstatistics based on conditional probabilistic concepts, which can take into account the existence of two largely separated time scales and an associated conditional entropy in such systems. We have recovered ordinary thermodynamics in the case when there are no temperature variations, and have systematically evaluated superstatistical corrections for systems with sharply peaked temperature variations. Moreover, we have discussed a generalization of Einstein's theory of fluctuations in conformity with a maximum entropy condition. We have also illustrated this on the very simple model of superstatistical Brownian particles.

We believe that our conditional entropy approach offers a useful basis for describing the macroscopic properties of a wide class of superstatistical complex systems in nonequilibrium stationary states. Also, the discussion can straightforwardly be generalized to systems in which there exist more than two separated scales, by the repeated use of Bayes' rule. In this way, one can then construct multiscale superstatistics and its corresponding thermodynamics, with possibly interesting resonance or interference properties, depending on the characteristic time scales in the systems.

## ACKNOWLEDGMENTS

This work was initiated during the International Summer School and Workshop on Complex Systems and Nonextensive Statistical Mechanics (July 31- August 8, 2006, The Abdus Salam International Centre for Theoretical Physics, Trieste, Italy). The authors thank ICTP for their hospitality and the stimulating atmosphere. The work of SA was supported in part by Grant-in-Aid for Scientific Research (B) of the Ministry of Education. CB acknowledges financial support by an EPSRC springboard fellowship. EGDC is indebted to the NSF for financial support by grant Phy-0501315.

———————————————————